# The Pattern of Solar Wind-Magnetosphere Interaction and Its Universality

Eugene Savov

Solar-Terrestrial Influences Laboratory, Bulgarian Academy of Sciences
Acad. G. Bonchev St., Block 3, Sofia 1113, Bulgaria

## Abstract

The multi-scale expansions and contractions of the Earth's magnetosphere explain fundamental issues of magnetic storm-substorm relationship. This magnetospheric behavior is in agreement with a model of 3D-spirally-faster-inward-oscillating dynamic fractal structure of the universe. This fractal like structure accounts for the found puzzling similarity between the near and most distant cosmos and for the calculated fractal dimensions of the observed distributions of galaxies. The introduced dynamic fractal dimension d(r) and formula for unifying force / F(r) = ma(r) = constant/[r]exp(d(r)) / present new fundamental framework for qualitative and quantitative modeling. Assessments and predictions based on the dynamic fractals are described.

## Introduction

The solar wind makes the Earth's magnetosphere to expand and contract as indicated by the expansion and contraction of the auroral oval. The dynamic pressure of the solar wind is balanced at inner (outer) lines in the expanding (contracting) magnetosphere and so auroral particle precipitation is delivered to lower (higher) latitudes (Savov, 1998). The multi-scale expansions and contractions of the magnetosphere account for fundamental issues of solar wind-magnetosphere interaction, e.g. the magnetic storm-substorm relationship (Savov, 1998). This magnetospheric behavior is in agreement with 3D-spirally-faster-inward contracting and expanding structure of one all-building unifying interaction that creates universal scale (distance) in way which accounts for what we see and described by laws of classical and quantum theories (Savov, 1991, 1993, 1998, 2002).

The study of the solar wind-magnetosphere interaction reveals a pattern of expanding and contracting magnetic configuration, which indicates three-dimensional (3D) spiral inward intensifying interaction (*Savov*, 1991, 1993, 1998). Understanding depends on what we see. What creates what we see is uncertain. This makes the understanding of nature very difficult. It is dramatically simplified by considering Mandelbrot's expanded notion of fractal, viewed as invariance under certain class of transforms (*Mandelbrot*, 1983). It says that the invariance under certain rules creates a fractal set, i.e. a set made of similar elements. The laws of all-building unifying interaction have to be scale independent and hence observer invariant. The scale dependence of the classical and quantum theories is likely to appear in the process of observation. Otherwise observer has to occupy a special place, which is somewhere between the deterministic and continuous classical and the probabilistic and discrete quantum realities. It is very likely that these two realities are generated in the process of perception of one poorly understood unified reality. A set of scale independent laws of the unifying interaction will generate a dynamic fractal like structure of reality. This 3D-spirally-faster-inward-moving structure of nature was first discovered in 1991 from an attempt for 3D closure of the magnetospheric currents (*Savov*, 1991).

It was later used to reveal the 3D-spiral-expansions and contractions of the Earth's magnetosphere, which account for fundamental issues of its behavior, e.g. the poorly understood field-aligned acceleration (*Savov*, 1991, 1998, 2002).

This paper demonstrates how the behavior of one dynamic fractal element of the 3D-spirally-faster-inward-moving basic matter (unifying interaction) accounts for the fractal dimensions of the observed galaxies distributions, for the inverse square laws and for the puzzling similarity between the near and most distant universe.

**Dynamic fractals**

Figure 1 shows the revealed from spacecraft observations dynamic fractal pattern of unifying interaction (basic matter). The 3D-spiral swirl of basic matter (P) attracts itself by moving 3D-spirally faster inward. Then it bifurcates and moves 3D-spirally-faster inward (P') and outward (P'') as shown by larger inward arrows in Fig. 1B. Later its outward motion decreases and it collapses on its source (Fig. 1C). It comes from and ends on its source. It is everything (*Savov*, 2002). Thus it creates universal scale, i.e. distance $r$ created from its structure. It depending on its ratio to the scale of the interaction that builds the process of observation creates what we see as electromagnetic interaction, space and time (*Savov*, 1993, 2002).

Dynamic fractals or in other words patterns of unifying interaction (Fig. 1) that have atomic sizes will show $div\mathbf{P} = 0$ and $div\mathbf{P} \neq 0$, respectively, far from their cores and for their cores, where $\mathbf{P}$ is the inward increasing the velocity of the basic matter that builds them. This allows obtaining Maxwell's equations of electromagnetic field (*Savov*, 2002). It is interesting to notice that $div\mathbf{P} \neq 0$ everywhere for the spiral pattern shown in Fig. 1, i.e. it creates non-zero flux through every surface. Thus it indicates its existence made of a finite number of dynamic fractal elements that have scale invariant properties.

Here the properties of the dynamic fractal shown in Fig. 1 will be studied in terms of introduced dynamic fractal dimension, which is also incorporated into a formula for the unifying force. The study of one element of the self-similarly evolving structure (Fig. 1) is equivalent to the study of the structure a whole. It will be shown that multi-scale appearances of the dynamic fractals (Fig. 1) account for the fractal dimension of galaxies distribution, the enigmatic similarity between the near and most distant cosmos, the Olber's paradox and the puzzling heat sources in the stellar and planetary cores.

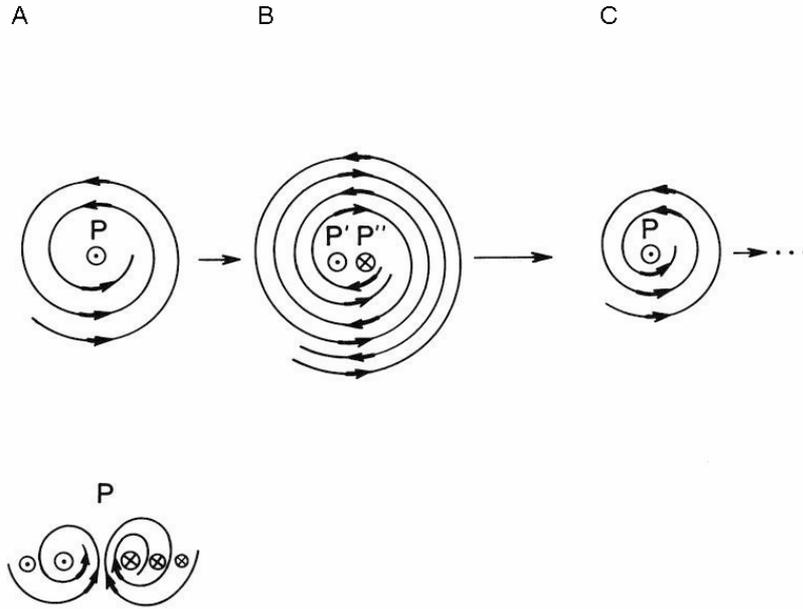

**Fig. 1.** This is a dynamic fractal of unifying interaction (basic matter) P that creates universal distance *r* (scale) from its core. It moves 3D-spirally-faster-inward as shown by larger inward arrows of its equatorial (above) and meridian (below) cross-sections (A), bounces back and starts to oscillate 3D-spirally-faster-inward (the meridian cross-section is not shown in (B) and (C)), finally it collapses on its source (C) and begins its cycle again. It oscillates 3D-spirally-faster-inward, accumulates its environment and ejects smaller similar ones that do the same. Thus it creates the finite sources of all-building unifying interaction in one self-consistent and complete picture of firework universe that goes through all annihilating cycles as it collapses on its source (C and A) is similarly born again (B), govern by similar laws of physics. (From Fig. 2 of *Savov*, 1993).

**Dynamic fractal dimension and unifying force**

We should learn to differentiate between what it is and how it is seen. The laws of the unifying interaction are in the logic of the dynamic fractals not in how they are seen. The logic, i.e. the self-consistent behavior, of the dynamic fractal shown in Fig. 1 explains everything in a dynamic fractal world that has 3D-spiral code. Everything evolves self-similarly and so we know ourselves in the mirror and study the rest of universe (*Savov*, 2002). Self-similar evolving gives rise to consciousness and makes nature accessible to mind.

Let the dynamic fractal pattern of interaction shown in Fig. 1 is built from one basic matter or in other words from one unifying interaction. The carriers of the unifying interaction show non-zero flux through any surface (Fig. 1) and so state their existence. One dynamic fractal element of Fig. 1A type creates universal scale (distance) *r* in the first moment of its creation. Then the number of fractal elements $N(r)$ that create the scale *r* is

$$N(r) = C_n r^0 = 1, \tag{1}$$

where $C_n$ is a counting constant corresponding to the accuracy of observation. The universal scale *r* means creation of a distance (difference) caused by one's existence. It can be seen as space, time, cosmic body or whatever exists, depending on its ratio to the scales created from the interactions that build the process of observation (*Savov*, 1993, 2002).

The faster inward 3D-spiral motion of the dynamic fractal attracts new similar elements as it self-similarly evolves. Then $N(r)$ increases and becomes finite in the smallest number of dimensions - 3D that allow this to happen as the dynamic fractal bounces off from itself (Fig. 1B). Hence one can write

$$N(r) = C_n r^3. \tag{2}$$

Then the 3D-spirally-faster-inward-moving, i.e. falling into itself dynamic fractal structure, bounces 3D-spirally outward and starts to oscillate 3D-spirally-faster-inward. The frequency of its oscillation is

$$f(r) = 1/T(r) \text{ and } T(r) = 2\pi r/v(r) \tag{3}$$

$v(r)$ = constant = $C_v$ defines the dynamic fractal by the velocity of its elements that create the distance $r$, $T(r)$ is the time it takes one fractal element to revolve at angle $2\pi$ into the fractal structure at distance $r$ from its core (Fig. 1). This is the distance between the center of the dynamic fractal and the fractal element moving around it. Hence

$$f(r) = C_v/2\pi r, \tag{4}$$

Equations (1-3) describe self-similar evolving dynamic fractal. They are generalized in

$$N(r) = C_n r^{d(r)} \tag{5}$$

where $d(r) \in [0, 3]$ is the dynamic fractal dimension, created by $N(r)$ number of dynamic fractals that build the distance (universal scale) $r$. In the beginning of dynamic fractal evolving $d(r) \geq 0$ and afterwards it increases to $d(r) \leq 3$ with the increase of the number of the 3D-spirally-faster-inward accumulated fractal elements that create the universal scale $r$ (Fig. 1A). The dynamic fractal dimension increases with the increase of the density of the dynamic fractals. Then the dynamic fractal starts to oscillate 3D-spirally-faster-inward with frequency $f(r)$ (Fig. 1B and Eqs. (3) and (4)). Then from Equations (3), (4) and (5) the number of fractal elements $N_o(r)$ creating the universal scale $r$ in the oscillating structure is

$$N_o(r) = 2f(r)N_c(r) = C_n \left[ v(r)/\pi \right] r^{d(r) - 1} \sim r^2 \tag{6}$$

The oscillation adds with twice the frequency $f(r)$ the number $N_c(r)$ of contracted fractal elements that create $r$. They fall 3D-spirally into themselves and bounce 3D-spirally outward doing this with frequency $f(r)$. The dynamic fractal element starts to oscillate 3D-spirally-faster-inward and so "comes to life" (Fig, 1B). The distance $r$ is made from 3D-spirally-faster-inward-oscillating dynamic fractal elements that are finite, discrete, unique and self-similarly evolving.

The volume of one 3D-spirally contracting and expanding (oscillating) dynamic fractal is $V_o^1(r) = V_o(r)/N_o(r) \sim 1/r^2$, where $V_o(r)$ = constant at the scales defining the structure built from $N_o(r)$ number of dynamic fractals. The 3D-spiral contractions and expansions of one self-similarly evolving fractal element (Fig. 1) create unifying force, which is proportional to its 3D-spirally contracting and expanding volume.

$F(r) = \sim V^I(r) = V(r)/N(r) \sim 1/r^{d(r)}$ and from Fig. 1 it is obtained

$F(r) = C_f/r^{d(r)}$ for $d(r) \in [0, 3]$ during the 3D-spiral collapse (Fig. 1A)   (7)

$F(r) = C_f/r^{d(r)}$, $d(r)= 2$ during the 3D-spiral oscillation (Fig. 1B)   (8)

$F(r_i) = C_i$ quantum force at the scales $r_i$ created from the oscillating structure (Fig. 1B)   (9)

$v^2(r_i)/r_i = const$, $r_i = const$ and $v^2(r_i)r_i = const$ and   (10)

$v(r_i)r_i = const$   (11)

$F(r) = ma(r)$   (12)

where $V(r) = $ constant defines the volume of the structure in which the dynamic fractal element belongs, $N(r) = C_n r^{d(r)}$ (Eq. (5)) is the number of dynamic fractal elements building that the volume $V(r)$, $V^I(r)$ is the volume of one fractal element whose 3D-spiral-faster-inward contractions and expansions create the unifying force $F(r)$, $d(r)$ is the dynamic fractal dimension, $C_f = V(r)/C_n = $ constant is unifying force constant, seen as unchanged during the interaction between the considered dynamic fractals that build the distance $r$, $m$, called mass, is the resistance to acceleration of a dynamic fractal, e.g. one seen as test body, merged into a 3D-spirally-faster-inward-moving dynamic fractal structure that creates $V^I(r)$ and $a(r)$ is the acceleration along $r$ created by the 3D-spiral dynamic fractal. Every motion is driven by outer and inner dynamic 3D-spiral fractals and so its trajectory shows, respectively, smaller and larger curvatures ($1/r$). The equations of the unifying force (7-9) show that the fractal dimension $d(r)$ of dynamic fractals increases up to 3 with their density increase (Eq. (7)). Afterwards the structure begins to oscillate 3D-spirally-faster-inward and its fractal dimension drops to 2 (Eq. (8)).

The oscillating structure (Fig.1 B) shows discrete (quantum) behavior at scales of its own (Eq. (9)). The dynamic fractal density increases creating smaller scales $r$ from 3D-spirally-faster-inward-oscillating cores (Fig. 1B). The increase of the density inward into the 3D-spirally-faster-inward-moving structure makes its coupling insides visible and accounts for the three dimensions of the visible bodies.

The increase of the distance $r$ decreases the number density of the dynamic fractals that create it. The initial dynamic fractals are always smaller in number than their outcomes. This leads to decrease of the fractal dimension below 3 in agreement with the obtained fractal dimensions about 1 to 2 from observed distributions of galaxies. This also solves the Olber's paradox, requiring bright night sky in a universe uniformly populated with average stars, whose number density increases as $r^3$ and light flux decreases as $1/r^2$, where $r$ is the distance from the observer. In the hierarchical dynamic fractal built firework universe the fractal dimension, i.e. the number density, of atoms, stars, galaxies etc. decreases with universal scale (distance) increase.

The initial (parent) dynamic fractals create unifying force according to Eq. (7). Afterwards the dynamic fractal starts to oscillate (Eq. (8)) and thus is ejected from the insides of the faster-inward-oscillating parent dynamic fractal (Fig. 16 in *Savov*, 2002). The result is a self-consistent and complete picture of firework universe made of 3D-spirally-faster-inward contracting and expanding, oscillating dynamic fractals that show discrete (quantum) behavior at scales of their own (Fig. 1B). Observer in the firework universe will drift within dynamic fractals seen as galaxies, stars and atoms at every distance. This is because the accessible universe is like a cloud of dynamic fractals seen as galaxies that move around their source. Similarly, but on much smaller scales an atmospheric cloud moves around the nucleus of the Earth from where its atoms have originated.

The universal scale, i.e. the distance $r$, created from the dynamic fractal core (Fig. 1) accounts for everything by appearing in different sizes (*Savov*, 1993, 2002). The smaller distances are $r_i$ are created from 3D-spirally-faster-inward-oscillating and hence seen as discrete (quantum) and denser dynamic fractals. The 3D-spirally-faster-inward-oscillating smaller scales are dense enough to emit and reflect much smaller similar ones seen as light. In this way 3D bodies seen as atomic matter that emits, reflects and engulfs light are created. The 3D-spiral-faster-inward contraction and expansion, oscillation makes the dynamic fractals finite and discrete (quantum) at scales of their own (Savov, 2002).

The scales of observation are created from the interaction between more than $10^{24}$ atoms that build one's brain ($N = 6.02 \times 10^{23}$ is Avogadro constant giving the number of molecules in one mole of any substance). Then the universal scale $r$ will show quantum behavior at atomic and smaller scales and continuous behavior at larger than the atomic scales. The discrete evolving of the larger scales will be integrated in the process of perception like movie frames. Single dynamic fractals (Fig. 1B), seen as light photons, will be discretely bent or engulfed by much larger dynamic fractals that build the atoms of a tin slit through which the light photons pass. This will create the observed bright and dark bands on a screen behind the slit. One of the big conundrums of the quantum world is so explained. The invariance and change of the dynamic fractals ($r$) at the scales of the interaction that builds the observation are perceived, respectively, as space and time.

Equations (10) and (11) applied for the solar system lead to the third and second Kepler's laws (*Savov*, 2002). Equation (11) applied to atomic nucleus and electron results in the principle of quantum uncertainty (*Savov*, 2002). Let us consider Equation (11) $v(r_i)r_i = const$. Then at atomic and smaller scales the accuracy of each measured coordinate is $\Delta x \sim r_i$ the distance created from a dynamic fractal core (Fig. 1B), which seen as atomic nucleus. At the this distance electron orbiting the atomic nucleus will change its velocity as $2v(r_i) = \Delta v_x(r)$. Hence from Eq. (11) $\Delta v_x(r) \times \Delta x = const$ and $\Delta p_x \times \Delta x = const$, which for $\Delta p_x = m\Delta v_x(r)$, where $m$ is the mass of the electron has the sense of the Heisenberg's uncertainty principle. The knowledge for the position $\Delta x \sim 0$ sets the particle in a fast rotation around the dynamic fractal core seen atomic nucleus. This smears the direction of the impulse of the particle. The knowledge of the velocity direction $v_x(r)$ runs out of the discrete atomic scales. This makes the position of the particle uncertain. The principle of uncertainty is just another confirmation of all-building dynamic fractals (Fig. 1).

In a universe made of 3D-spiral dynamic fractals (Fig. 1) everything is curved and discrete (quantum) at its finite scales. Hence the dynamic fractal dimension and the unifying force constant take different values for different dynamic fractals, particularly when they belong to different hierarchies of origin. The cores of the dynamic fractals (Fig. 1B) are made of strong (stable), discrete (quantum) 3D-spirally-faster-inward-oscillating interactions.

Everything is discrete like movie frames in a world made of 3D-spirally-faster-inward-oscillating dynamic fractal elements, i.e. patterns of unifying interaction, ordered along a hierarchy of origin to create what we see as galactic nuclei, stars, planets, planetary type moons, atoms and light (Savov, 2002). The dynamic fractals contract and expand, oscillate faster–inward, and so eject smaller similar ones from their cores. That is why we see light, i.e. much smaller dynamic fractals of Fig. 1B type, emitted from the agitated new atomic matter created in the stellar nuclei.

The discrete values of the dynamic fractal dimension and the unifying force constant in the equation of the unifying force (Eq. (7)) can be seen as constants at the accuracy of observation, made from the interactions that build it. We see interacting oscillating dynamic fractals (Eq. (8)) at the atomic scales of observations from which the gravitational constant ($G$) and the absolute permittivity of the medium ($\varepsilon$) are obtained. Then the equation of the unifying force (Eq. (8)) turns into the well known Newton's law of gravitational attraction and Coulomb's law for coupling electric charges. The nature of the unifying force captured in Equation (8) shows that the gravitational force between two spheres will depend on their material and temperature. It is a matter of accuracy of measurement to reveal these dependencies. Nothing is the world of dynamic fractals is constant although some interactions will show constant parameters (like electric charge, mass, etc.) at the scales of observation. This creates the notion of fundamental constants.

The unifying force in the dense dynamic fractal regions, e.g. at atomic scales, will show discrete properties at the scales of observation. Then Equation (8) takes the form of $F(r_i) = C_i = $ const, where $r_i$ are the discrete distances between the cores of two closely (strongly) coupling 3D-spirally-faster-inward-oscillating dynamic fractals. In case of dynamic fractals seen as atomic nucleus and electron the distances $r_i$ between their cores correspond to the quantum levels of the electron. So the equation of unifying interaction accounts for the *observed* quantum properties of the micro world. **The dynamic fractal structure of nature is quantum everywhere at scales of its own.** The ratio between its scales and that created by the process of observation creates what looks like continuously moving planets and discrete energy levels (distances $r_i$) of electrons. The dynamic fractal dimension evolves from 1 to 3 (Fig. 1A) and then becomes 2 when the dynamic fractal "comes to life" by appearance of 3D-spiral-faster-inward oscillation (Fig. 1B). The fractal dimension of the oscillating fractals decreases (increases) with $r$ increase (decrease) as the secondary fractal elements are driven in the space created from the initial ones (or come closer).

## Qualitative and quantitative assessments

The bulk of the Sun is made mostly of Fe, Ni, O, Si, S (in decreasing order) and at about 1% atomic abundance level there are He, C, Ne, Ca and Cr (*Manuel and Hwaung*, 1983). These findings "suggest that fusion of hydrogen is probably not the Sun's primary energy source" (*Manuel and Hwaung*, 1983). These very confusing results are simply explained with creation of new atomic nuclei from the huge stellar type nucleus of the sun. The lighter nuclei protons, electrons and He++ have smaller interaction cross-sections. So they much easily find their way upward into the solar wind. The heavier nuclei /Fe, Ni, O, Si, S/ remain closer to their source and create the atomic matter build bulk of iron sun. The creation of new atomic matter in the stellar type nuclei of stars, planets and planetary like moons simply accounts for their hot insides, volcanic activity, hot spots and for the puzzling inner heat source of Jupiter and the other gas giant planets.

"When elemental abundances in the photosphere are corrected for this mass separation, the most abundant elements inside the Sun are Fe, Ni, O, Si, S, Mg and Ca" and these elements comprise ≈ 99% of the ordinary meteorites (*Manuel and Kamat*, 2005). This finding suggests that meteorites are remnants from exploding stars and most likely from exploding small cooled star that was a planet, which moved around the Sun, somewhere between the orbits of Mars and Jupiter. Its explosion created the asteroid belt and probably led to the extinction of the dinosaurs (*Savov*, 2002). The heavier atomic nuclei remain closer to their sources – stellar type nuclei. Latter most of the heavier nuclei are annihilated by their collapsing sources in supernova events. So the explosive beginnings and ends of stars in the firework universe (*Savov*, 2002) populate the cosmic space mostly with light elements H and He, found in the account for more than 95% of the solar wind. That is why we currently believe in nucleosynthesis appearing in atomic matter collapsing into stars. The stars repel in the dense stellar populations because. They act like huge atomic nuclei because of their 3D-spirally-faster-inward-oscillating structure shown in Fig. 1B.

The transition from dynamic fractal dimension $d(r) = 2$ to $d(r) = 3$ means collapse of the oscillating structure at considered dynamic fractal created scales. This is seen as appearance of antimatter behavior and fractal density increase (Fig. 1B turns in Fig. 1C). The increase of the dynamic fractal number density $N(r)$ creates fractal dimension 3. This is the case of atomic matter that fills a volume of about 1 $m^3$. The number of the atoms in 1 $m^3$ of non transparent (dense) material is about $10^{27}$. This number is comparable or larger than the number of the galaxies in the accessible universe. The fractal dimension will decrease toward larger scales due to creation of $r$ by initial (larger) dynamic fractals. Hence the fractal dimensions of galaxies distribution is likely to be between about 1 and 2, implying oscillating dynamic fractals mixed by much larger (initial) ones of their parent structure.

All phases of Fig. 1 /A, B and C/ coexist at certain finite scale as they apply for the different $r$ they create. In the case of solar system $F(r) \sim 1/r^2$ near the Sun where the planets move in the oscillating dynamic fractal whose core is seen as the sun. The accretion of fractal elements from the parent Galactic dynamic fractal structure leads to adding of a new force $F(r) \sim 1/r^{d(r)}$, where $d(r) < 2$. This directly accounts for the puzzling force that tugs the receding Pioneer spacecraft sunward. The measurement of this sunward force along the path of the spacecraft will lead to obtaining of its fractal dimension.

The formula of the unifying force applies for coupling dynamic fractal elements. They interact by fitting their universal scales (*Savov*, 2002). Dynamic fractal elements of the same hierarchy of origin interact and build structures. For example, we see coupling chemical elements, multiple stars and clusters of galaxies. The fractal dimension, the unifying force constant $C_f$ and also the structure (the origin) of the distance $r$ vary more or less in the formula for the unifying force, when it is applied along different scales. The unifying force describes the self-similarly evolving dynamic fractal shown in Fig. 1. The regularities obtained from one dynamic fractal element (Fig. 1) apply for all elements of the dynamic fractal structure.

The all-building unifying interaction has dynamic fractal structure like that shown in Fig. 1. It increases 3D-spirally-faster-inward and becomes finite in 3D through initiation of its 3D-spiral-faster-inward contraction and expansion, i.e. oscillation. Then $N(r) = N_o(r)$ and from Eqs. (6) the dynamic fractal dimension is $d(r) = 2$. The increase of the frequency $f(r)$ of dynamic fractals oscillation inward with their density increase (Fig. 1) makes them show up as dense 3D objects that can emit, reflect and engulf much smaller ones seen as light.

The faster-inward 3D-spiral motion of basic matter (i.e. unifying interaction, fractal elements) brings the bodies (i.e. the dynamic fractals) together (Fig. 1A). The outward 3D-spiral motions keep them apart (Fig. 1B). This explains the observed clustering of matter at cosmic and atomic scales. The increase of the 3D-spiral dynamic fractals density increases the fractal dimension. This accounts for the observed 3D dimensions of the bodies because they are dense enough to reflect or emit much smaller ones seen as light. The inward 3D-spiral swirl of basic matter prevails. It increases relatively to the outward 3D-spiral swirl further from their center. The puzzling force that tugs Pioneer spacecraft sunward (*Turyshev et al.*, 2005) is explained with the 3D-spiral code of the unifying interaction considered at the scales of the solar system (*Savov*, 2005a). Simply the inward 3D-spiral vortex of basic matter is increasingly stronger than the outward one further from the core of the dynamic fractal (Fig. 1B). The inward vortex it attracts dynamic fractal elements from its parent galactic structure. Then the inward swirl of basic matter tugs the spacecraft sunward (inward) as it flies away from the sun. Hence the unifying constant and the dynamic fractal dimension change in the equation of the unifying force (Eq. (8)). The large distance from the Sun is created from dynamic fractal elements of the parent Galactic structure. These elements wind up into the inward 3D-spiral swirl in which core we see the sun. So the observed confusing sunward attraction is created.

The unifying force is defined in the dynamic fractal structure shown in Fig. 1. Its finite elements create finite and discrete universal scales that correspond to what we see as space, time, bodies, light and whatever exists (*Savov*, 2002). Nothing is completely equivalent to anything else because every two dynamic fractal elements differ in a finite scale, created from their unique dynamic fractal structures (universal scales). The elements can look equivalent at the finite accuracy of observation and are described with the classical and quantum theories (*Savov*, 2002).

The dynamic fractal creates universal scale $r << r_s$, where $r_s$ is the scale of its source. Then Equations (7) written for its source suggests appearance of strong unifying (all-building) interaction that 3D-spirally collapses into itself, starts to oscillate 3D-spirally-faster-inward and thus becomes a source for similar dynamic fractals, i.e. finite patterns of unifying interaction, that do the same and so on. The result is a cyclic firework universe, made of finite self-similarly 3D-spirally-faster-inward-oscillating dynamic fractals. They create the fundamental framework for complete understanding in which phenomena are simply explained directly in the terms of their origin (structure, source). These multi-scale dynamic fractals, called protobodies, account for what we see as space, time, cosmic bodies and everything (*Savov*, 2002). Nature is made of multi-scale oscillators, whose frequencies increase inward. They self-organize (synchronize) and unfold the unifying interaction that builds them (*Savov*, 2002, 2005b). The 3D-spirally-faster-inward-oscillating structure of reality ejects similar dynamic fractals from its insides. This universal drive for unfolding is the force of self-organization and life.

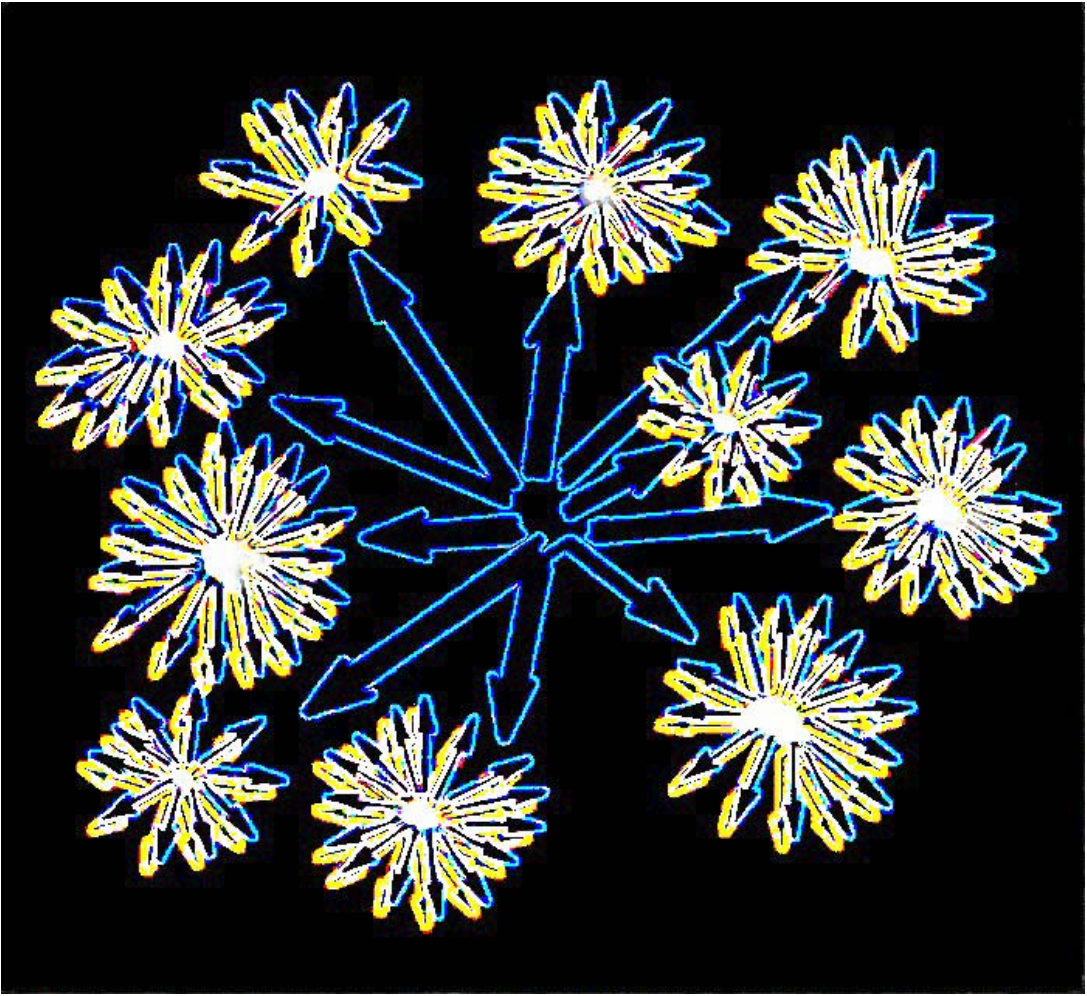

**Fig. 2.** It 3D-spirally-faster-inward-oscillates and ejects smaller similar ones like fireworks from its insides (Fig. 1). The created multi-scale 3D-spiral swirls of basic matter (unifying interaction) interact and build the observed clumpy structure at every scale. Smaller 3D-spiral swirls are ejected from the insides of finite similar larger ones. In this way very small 3D-spiral swirls seen as light are released. We see only the dense insides of the 3D-spiral dynamic fractals (Fig. 1) that emit or reflect light.

The number of dynamic fractal elements, seen as galaxies, will obey Eqs. (5-8). Then the oscillating fractals observed as galaxies will show dynamic fractal dimension $d(r)$ smaller than 3 depending on the density of their distributions. The dynamic fractal dimension connects the number of dynamic fractals $N(r)$ with the distance $r$ they create - $N(r) = C_n r^{d(r)}$, where $C_n$ is the counting constant. It corresponds to the calculated fractal dimension D of galaxies, which in the simplest case shows that the number of galaxies within given volume is proportional to $R^D$, where R is the linear scale of the volume and D is the fractal dimension of galaxies distribution. The calculated fractal dimensions of about 1 to 2 of galaxies distributions (*Labini, et al.*; *Joyce et al.* 2005; *Celerier and Thieberger*, 2005 and the references therein) suggest, respectively, less or more strongly interaction galaxies that drift in the space created from their sources - Great Attractor (*Dressler*, 1991) type objects.

The number of galaxies around given galaxy within a distance R has to be proportional to $R^D$, where D is fractal dimension between 0 and 3. The uniform distribution requires D = 3 and D < 3 implies fractal. The observed fractal dimensions of galaxies distribution is about 2.1 and "the Universe is fractal on scales up to 300 million light years" (*Labini, et al.*). ""Our tests show that the Universe never becomes homogeneous in the available galaxy samples," says Sylos Labini, who began his work while in Pietronero's team. "It remains hierarchically clustered. It remains fractal." (*Labini, et al.*)." This is what is expected from the hierarchical dynamic fractals that 3D-spirally-faster-inward-oscillate and eject smaller similar ones. The dynamic fractal ejects smaller similar elements from its insides and drives them around. Only the denser insides of the hierarchical 3D-spirally-faster-inward-oscillating structure are observed and the rest accounts for the nature of dark matter. The invisible part of discovered hierarchical dynamic fractal like structure of the universe explains the stability of galactic clusters and reveals the uncertain nature of dark matter. The obtained from observations fractal dimension about 2.1 is in agreement with the theoretically expected smaller than 3 one, for weakly interacting galactic size dynamic fractals (Eqs. (6-8)).

The galaxies organize in dense regions and voids (e.g. *Einasto et al.*, 1997). In this way their 3D-spirally-faster-inward-oscillating structures come closer and expand to create the voids around their coupling inner regions. Two coupling structures of galaxies will rather repel than merge like two forms of life. Anyway the constant mixing of the galactic structures, as indicated by their smaller than 3 fractal dimensions, by the initial dynamic fractals will make the secondary structures grow in a patterns characterized by connected dense regions and voids. Similar process of universal pattern formation works at atomic scales, mixed by the dynamic much huger fractals of the sources of atoms. The cores of these sources are in the centres of stars, planets and planetary like moons and create their stellar type nuclei. The planets and their planet like moons are small cooled stars.

The dynamic fractals, i.e. the bodies, of the firework universe have explosive beginnings and ends, created as they are ejected from their 3D-spirally-faster-inward-oscillating sources and later collapse on them. Some of the explosive ends are seen as supernovas. Then the atomic matter shell of the star is annihilated as its source collapses at atomic scales. Then the core of 3D-spirlly-faster-inward-moving dynamic fractal, seen as a star, is observed like a fast rotating very dense object, known as pulsar and considered as neutron star in the incomplete picture of the big bang universe. The explosive beginnings and ends of the stars populate the cosmic space predominantly with light dynamic fractal elements seen as H and He. The heavier chemical elements are more strongly coupled to their sources and finally annihilated by them.

The observed constant speed light is directly obtained from Fig. 1B, when it is considered as agitated atomic nuclei. It 3D-spirally-oscillates-faster-inward, over spins and ejects much smaller similar ones seen as light. The size of the pattern of unifying interaction is defined by the velocity of the unifying interaction carriers at distance $r$ from the center of the pattern. Then the velocity $v(r_i) = const$ from Eq. (10) in the cores of the dynamic fractals, seen as atomic matter. The velocity of elements seen light ejected at distance $r_i$ from the center of a dynamic fractal, seen as atom, has to be constant if the atoms are stable to create the observer and the process of observation. The velocity of dynamic fractal elements has to be constant if its sources are to be stable at the scales of observation while creating the observer and what he finds in his mind. This creates the observed constant speed of light, which will move unobstructed in the much less dense dynamic fractals seen as vacuum. Light comes from agitation of the 3D-spirally-faster-inward-oscillating atomic cores.

The 3D-spirally-faster-inward-oscillating dynamic fractals, seen as atomic matter, will show discrete structure during high energy collision studies. This structure is discussed as protons, neutrons, quarks and elementary particles. The self-consistent picture of firework universe that accounts for many puzzling observations from the ambient and most distant space (*Savov*, 2002) cannot be obtained from structure matter inferred from collision studies. The latter suggest inverse problem that requires extra assumptions. The fundamental dynamic fractals oscillate and so eject similar ones. The created multi-scale dynamic fractals interact and build the observed clumpiness of matter at cosmic and atomic scales. Thus a self-consistent, complete, direct and simple picture of firework universe is created. The incomplete picture of the big bang created universe starts from singularity allowing point in which particles are born from mysterious matter-antimatter asymmetry. The initiation of density fluctuations and structure buildup from flying away from each other particles in 4D space-time is uncertain.

The 3D-spiral-faster-inward-oscillating elements of the fractal like universe, seen as planets and electrons move into much larger similar 3D-spiral parent structures, seen as stars and atomic nuclei. Hence planets and electrons cannot move around at arbitrary distances from the cores of 3D-spiral swirls, seen as stars and atomic nuclei, because the 3D-spirally-faster-inward-oscillating pattern of the unifying interaction is discrete. **The result is electrons abiding at quantum levels and planets found at distances from the Sun that correspond to Fibonacci numbers belonging to a spiral curve.**

The fundamental elements of nature are 3D-spirally-faster-inward-moving dynamic fractals created from one basic matter or in other words from one all-building unifying interaction. They are the finite, curved, discrete, unique, self-similarly evolving fundamental bricks of reality. Their size is hierarchically created along their origin. The smaller fundamental dynamic fractals are ejected from the oscillating insides of finite similar larger ones. We see only the insides of these fractals which are dense enough to reflect, emit or engulf light. The huger fractals, which are the sources of the atomic matter and their sources, create what we see as cosmic space and account for the uncertain nature of dark matter and for the cosmic structure buildup. The coupling of galactic size dynamic fractals creates the stability of galactic clusters in the same way the stability atomic structures is created at much smaller scales.

Every body is ejected from the insides of its source and is driven around by its outer 3D-spirally moving structure. That is why we see stars moving around galactic nuclei, planets orbiting stars, atomic matter moving around the nuclei of stars and planets and so accounting for the observed rotation of these space bodies. Every body moves around its source. All bodies move around the 3D-spirally-faster-inward-oscillating source of the universe that ejected and ejects smaller ones that do the same and so on. In this way the fractal like cyclic firework universe unfolds. Later it collapses due to the decrease of its 3D-spiral outward motion and is similarly born again, govern by similar laws of physics (Figs. 1 and 2).

In the beginning of the current firework universe multi-scale bright blue stars were created. The smaller moved around their sources the larger ones all moving around their source – the hyper huge nucleus, which is in the core of the most initial 3D-spirally-faster-inward moving dynamic fractal. The cooling this universe as whole created the cosmic background and its structure. The smaller stars cooled faster and become planets. The explosive end of a planet which moved around the Sun somewhere between the orbits of Mars and Jupiter created the asteroid belt (*Savov*, 2002).

The creation of atomic matter in the centers of stars and planets keeps their interiors hot. This leads to increase of the radius of the Earth and to slowing of its rotation as the Earth ages. The latter is in agreement with experimental studies (e.g. *Williams*, 2002). Similarly on much smaller scale the increase of the radius of the aging light coming from the distant galaxies creates the cosmic red shift (*Savov*, 2002). The dynamic fractal pattern shown in Fig. 1 is inferred from an attempt for 3D-clouser of the magnetospheric currents in 1991 (*Savov*, 1991). This pattern is confirmed with the multi-scale expansions and contractions of the Earth's magnetosphere and many confusing observations (*Savov*, 1998, 2002). Its multi-scale appearances account for the similarity between the near and distant universe and the fractal dimensions of galaxy distributions. Nature is made of dynamic 3D-spirally-faster-inward-oscillating fractals of basic matter (unifying interaction). The code of the universe is 3D spiral (*Savov*, 2002). The dynamic fractals contract and expand and thus eject smaller similar ones. In this way the finite similar source of reality are created. They account for the observed similarity across the scales and for whatever exists.

Every body is ejected and moves around its source. The observed universe is like a cosmic cloud of galaxies and sources of the galactic nuclei, which all move around their even much huger source – the core of much larger similar 3D-spirally-faster-inward-oscillating structure (Fig. 1) from which they were ejected. Observer in this cosmic cloud will see similar galaxies, structures of galaxies and light and heavy elements at very distance. The surprising observations galaxies, galactic structures and heavy elements in the distant universe (e.g., *Laurence et al.* 2005; *Ouchi et al.* 2004; *Mullis et al.* 2005; *Elston et al.*, 1994) are in agreement with this prediction of firework universe made of dynamic fractals having 3D-spiral code shown in Fig. 1.

One basic matter 3D-spirally falls into itself and bounces 3D-spirally outward (Fig. 1). In this way it creates the finite 3D-spirally-faster-inward-oscillating sources of reality. Their multi-scale 3D-spiral contractions and expansions make them self-defined. The created universal self-definiteness leads to appearance of consciousness that tries to define, to understand its source - the mind and the rest of the universe.

**Conclusion**

The dynamic fractals of unifying interaction and the unifying force (Eqs. (7-9)) create new fundamental framework for qualitative and quantitative modeling. The essential results are summarized below.

1) The calculated from observations fractal dimensions of galaxies distributions are in agreement with their dynamic fractal dimensions obtained in terms of 3D-spirally-faster-inward-oscillating fractals belonging to a revealed hierarchy of origin.

2) The unifying force works in hierarchically ordered along its origin dynamic fractal structure, created from finite self-similarly evolving 3D-spiral elements (Eqs. (7-9)).

3) The unifying force generates the well known inverse square laws of gravity and electrostatic interactions, when considered in the context, i.e. at the scales, from which these interactions were inferred.

4) The Pioneer spacecraft anomalous sunward acceleration is created from relative increase of the inward (sunward) element of the oscillating dynamic fractal whose core is seen as the Sun.

5) The observed constant speed light (a), the cosmic redshift (b) and the increase of the length of the day as Earth ages (c) are explained, respectively, with the stability (a) and the unfolding (b) and (c) of the dynamic fractal structures that build these phenomena.

6) It is shown that observer in dynamic fractals created firework universe will see galaxies, galactic structures, light and heavy chemical elements in the near and most distant space.

7) Variations in the gravitational constant depending on the material and the temperature of gravitationally coupling spheres are predicted.

The laws of the all-building unifying interaction are similar across the scales of nature and create similar outcomes. The universe is made of self-similarly evolving dynamic fractal elements having 3D-spiral code. The universe is round filled with round bodies ejected from the insides of similar finite larger ones. The bodies move around the cores of their sources, driven by their outer structure. That is why we see stars moving around galactic nuclei, planets orbiting stars and atoms moving around the nuclei of stars and planets. The observed rotation of space bodies is thus directly and simply explained.

The 3D-spirally-faster-inward contracting and expanding dynamic fractal (Fig. 1) creates one unifying interaction. It evolves self-similarly creating all-explaining hierarchy of origin (structure). Its multi-scale similar outcomes create what we see and describe in the laws and the principles of the classical and quantum theories (Savov, 2002). The laws of the unifying interaction are independent on how we see the coupling bodies. They are based on the properties their dynamic fractal shown in Fig. 1. Every body shows similar cyclic behavior at scales of its own. It comes from its expanding source. Afterwards it ends with its final contraction enjoying a state of existence (life) during its oscillation at its scales. The laws of the unifying interaction are captured in the equation of the unifying force (Eq. (7)). It applies for the all-building dynamic fractals considered at the scales of the studied phenomena. The calculations which do not follow the found dynamic fractal structure of nature run out of its scales and make no physical sense (*Savov*, 2002).

Reality is made of one all-building, scale-independent unifying interaction. Its dynamic fractal pattern accounts for what we see and describe in the laws of the classical and quantum physics (*Savov*, 2002). The parameters of the unifying force (Eqs. (7-9)) will be obtained for dynamic fractals building different phenomena in future research projects. The experimental data will be considered in terms of its revealed dynamic fractal structure. Fundamentally nature is extremely simple. It has a dynamic 3D-spiral code.

# References

Celerier, M.-N., and Thieberger, R. (2005), Fractal dimensions of the galaxy distribution varying by steps? http://www.arxiv.org/abs/astro-ph/0504442

Dressler, A. (1991) The Great Attractor: Do galaxies trace the large-scale mass distribution? *Nature*, **350**, 391-397.

Einasto, J, Einasto, M., Gottlöber, S., Müller, V., Saar, V., Starobinsky, A. A., Tago, E., Tucker, D., Andernach, H., and Frisch, P. (1997) A 120-Mpc periodicity in the three-dimensional distribution of galaxy superclusters, *Nature*, **385**, 139-141.

Elston, R., Thompson, K. L , and Hill, G. J. (1994) Detection of strong iron emission from quasars at redshift z>3, *Nature*, **367**, 250-251.

Joyce, M., Labini, F. S., Gabrielli, A., Montuori, M., Pietronero, L. (2005) Basic properties of galaxy clustering in the light of recent results from the Sloan Digital Sky Survey, http://arxiv.org/abs/astro-ph/0501583

Labini, S. et al., Fractured Universe, http://www.fortunecity.com/emachines/e11/86/fractured.html

Laurence, E., Bunker, A., Stanway, E., Lacy, M., Ellis, R., Doherty, M. (2005) Spitzer Imaging of i'-drop Galaxies: Old Stars at z~6, http://uk.arxiv.org/abs/astro-ph/0502385

Mandelbrot, B. B., On fractal geometry, and a few of the mathematical questions it has raised, in: *Proceedings of the International Congress of Mathematicians*, Warszawa, August 16-24, 1983, Z. Ciesielski and C. Olech, eds., PWN - Polish Scientific Publishers - Warszawa and Elsevier Science Publishers, B.V., P.O. Box 1991, 1000 BZ Amsterdam, The Netherlands, Vol.2, pp. 1661-1675.

Manuel, O.K., and Hwaung, G., Solar Abundances of the Elements, Meteoritics, **18**, No. 3 (1983) pp. 209-222 of http://web.umr.edu/~om/archive/SolarAbundances.pdf

Manuel and Kamat, "ISOTOPES TELL SUN'S ORIGIN AND OPERATION", submitted website poster presentation at The First Crisis In Cosmology Conference (CCC-I), Monção, Portugal, June 23-25, 2005. http://www.cosmology.info/2005conference/

Mullis, C.R., Rosati, P., Lamer, G., Boehringer, H., Schwope, A., Schuecker P., and Fassbender, R. (2005) Discovery of an X-ray-Luminous Galaxy Cluster at z=1.4, *Astrophysical Journal Letters*, 2005, ApJ, 623, L85. http://arxiv.org/abs/astro-ph/0503004

Ouchi, M., Shimasaku, K., Akiyama, M., Sekiguchi, K., Furusawa, H., Okamura, S., Kashikawa, N., Iye, M., Kodama, T., Saito, T., Sasaki, T., Simpson, C., Takata, T., Yamada, T., Yamanoi, H., Yoshida, M., Yoshida, M. (2004) The Discovery of Primeval Large-Scale Structures with Forming Clusters at Redshift 6, *Astrophys. J.* 620 (2005) L1-L4 http://arxiv.org/abs/astro-ph/0412648

Savov, E. P., On the closure of the magnetospheric currents, *Compt. rend. Acad. bulg. Sci.*, **44**, No. 5, 1991, 29-32.

Savov, E. P., On the character of solar-terrestrial interactions, *Bulg. Geophys. J.*, **19**, No. 4, 1993, 57-63.

Savov, E. P., On the magnetic storm-substorm relationship, *Bulg. Geophys. J.*, **24**, Nos. 3/4, 1998, 39-49.

Savov, E., *Theory of Interaction the Simplest Explanation of Everything*, Geones Books, 2002.

Savov, E., (2005a), Existing and unique firework universe and its 3D-spiral code,

*1st CRISIS IN COSMOLOGYCONFERENCE (CCC–I)*, Program, Moncao, Portugal, June 23 – 25, 2005. http://www.cosmology.info/2005conference/program.pdf

Savov, E., (2005b), Theory of Interaction Generated from Revealed Universality of Solar Wind Magnetosphere Coupling, Abstract, *5th Understanding Complex Systems Symposium*, University of Illinois at Urbana-Champaign, May 16-19, 2005. http://www.how-why.com/ucs2005/abstracts/Savov.html

Turyshev, S. G, Nieto, M. M., and Anderson, J. D., (2005) Study of the Pioneer Anomaly: A Problem Set, http://arxiv.org/abs/physics/0502123

Williams, G. E. (2002) Geological constraints on the Precambrian history of Earth's rotation and the Moon's orbit, *Rev. Geophys.*, **38**, No.1, 37-59.